\documentclass[american,english]{appolb}
\usepackage[utf8]{inputenc}
\usepackage{amsmath}
\usepackage{amssymb}
\usepackage{graphicx}

\makeatletter

\providecommand{\tabularnewline}{\\}
\newcommand{\lyxdot}{.}

\usepackage{epsfig}\usepackage{color}\usepackage{braket}

\makeatother

\usepackage{babel}
\begin{document}

\title{Detailed study of the quark-antiquark flux tubes and flux tube recombination%
\thanks{Presented at Excited QCD 2012, 6-12 May 2012, Peniche(Portugal)%
}}

\author{Marco Cardoso, Pedro Bicudo, Nuno Cardoso \address{CFTP, Instituto Superior Técnico} }
\maketitle
\begin{abstract}
In this work we compute the color fields in the mediator plane between
a static quark and a static antiquark using quenched lattice QCD.
In special, we see the effect of the quark-antiquark distance on the
flux tube. To obtain this results an improved multihit technique is
developed and an extend smearing technique is used. Then, we also
discuss the flux-tubes in a system composed of two quarks and two
antiquarks. The ground and first excited states fields are studied
for different dispositions of the system.
\end{abstract}
\PACS{12.38.Gc, 11.15.Ha}

\section{Introduction}

This work is divided in two parts. In the first one we will study
an isolated fundamental flux-tube in a quark-antiquark system. In
the second one we will study the fields a system made of two quarks
and two antiquarks, where we can observe the interaction between two
fundamental flux tubes.

\section{Chromo-fields computation}

To calculate the gauge invariant squared chromoelectric and chromomagnetic
fields in the lattice, we only have to calculate the correlation of
the Wilson loop operator $W$ which in large euclidean time limit
$t\rightarrow\infty$ should describe the static system, with the
plaquettes corresponding to each one of the fields. Concretely, we
have:

\begin{eqnarray}
\langle E_{i}^{2}\rangle & = & \langle P_{0i}\rangle-\frac{\langle W\, P_{0i}\rangle}{\langle W\rangle}\label{eq:plaqE}\\
\langle B_{i}^{2}\rangle & = & \frac{\langle W\, P_{jk}\rangle}{\langle W\rangle}-\langle P_{jk}\rangle\label{eq:plaqB}
\end{eqnarray}
where $P_{\mu\nu}=1-\frac{1}{3}\mbox{Tr}[U_{\mu}(\mathbf{s})U_{\nu}(\mathbf{s}+\boldsymbol{\mu})U_{\mu}^{\dagger}(\mathbf{s}+\boldsymbol{\nu})U_{\nu}^{\dagger}(\mathbf{s})]$.
The spatial indices $j$ and $k$ complement index $i$. The Lagrangian
density is given by $\mathcal{L}=\frac{1}{2}(E^{2}-B^{2})$.

\section{Fundamental flux-tube}

For a simple, quark-antiquark singlet, the Wilson Loop operator $W$
describes the system. We compute the chromo-fields in the mediator
plane between the quark and the antiquark, using 1100 pure gauge $32^{4}$
configurations with $\beta=6.0$ . In order to improve the signal
to noise ratio, we use two techniques: an improved version of the
multihit \cite{Parisi:1983hm} and an extended spatial smearing technique.

\subsection{Signal to noise ratio improvement}

In the multihit method we replace each temporal link by its thermal
average, with it's first neighbors fixed, that is $U_{4}\rightarrow\overline{U}_{4}=\frac{\int dU_{4}\, U_{4}\, e^{\beta\mbox{Tr}[U_{4}F^{\dagger}]}}{\int dU_{4}\, e^{\beta\mbox{Tr}[U_{4}F^{\dagger}]}}$.

We generalize this method by replacing each temporal link
by it's thermal average with the $N^{th}$ neighbors fixed, that is:

\begin{equation}
U_{4}\rightarrow\overline{U}_{4}=\frac{\int[\mathcal{D}U]_{\Omega}\, U_{4}\, e^{\beta\sum_{\mu\mathbf{s}}\mbox{Tr}[U_{\mu}(\mathbf{s})F_{\mu}^{\dagger}(\mathbf{\mathbf{s}})]}}{\int[\mathcal{D}U]_{\Omega}\, e^{\beta\sum_{\mu\mathbf{s}}\mbox{Tr}[U_{\mu}(\mathbf{s})F_{\mu}^{\dagger}]}}
\end{equation}

In this way we have an error reduction greater than the one of multihit.

\begin{figure}
\begin{centering}
\begin{tabular}{ccc}
\includegraphics[bb=130bp 100bp 280bp 280bp,clip,width=0.16\textwidth]{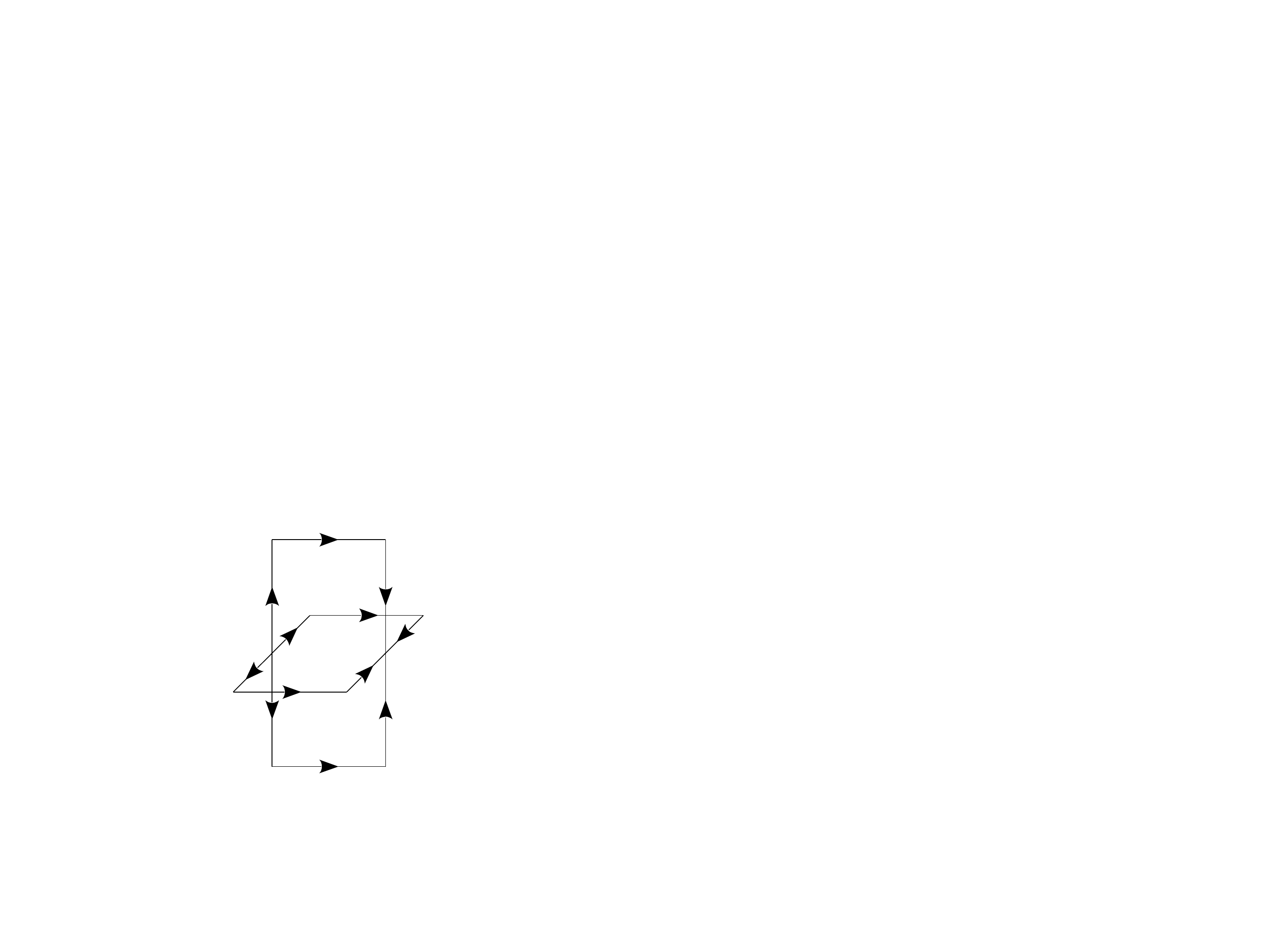} &  \includegraphics[bb=320bp 80bp 480bp 300bp,clip,width=0.18\textwidth]{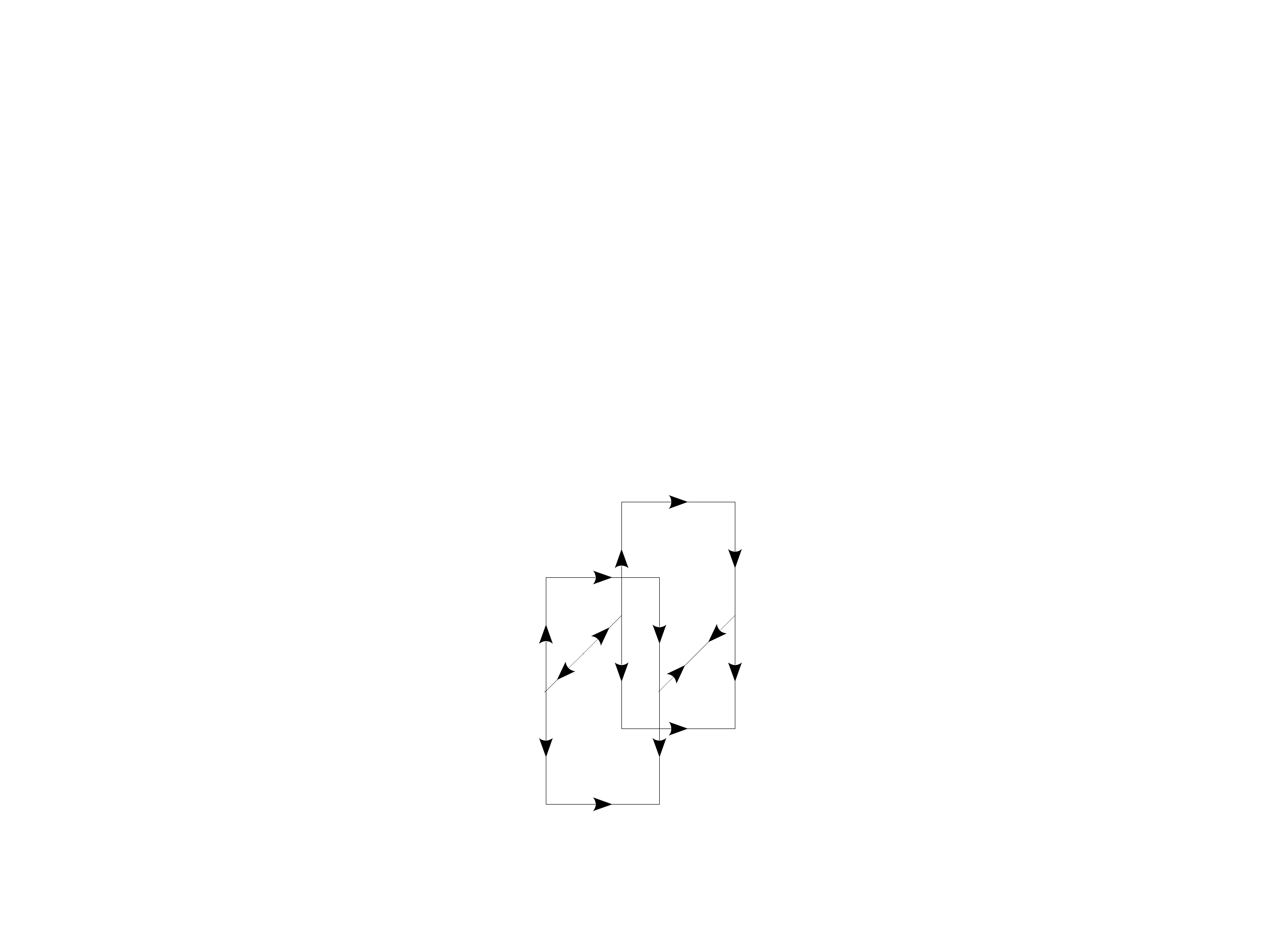} & \includegraphics[bb=510bp 30bp 700bp 340bp,clip,width=0.21\textwidth]{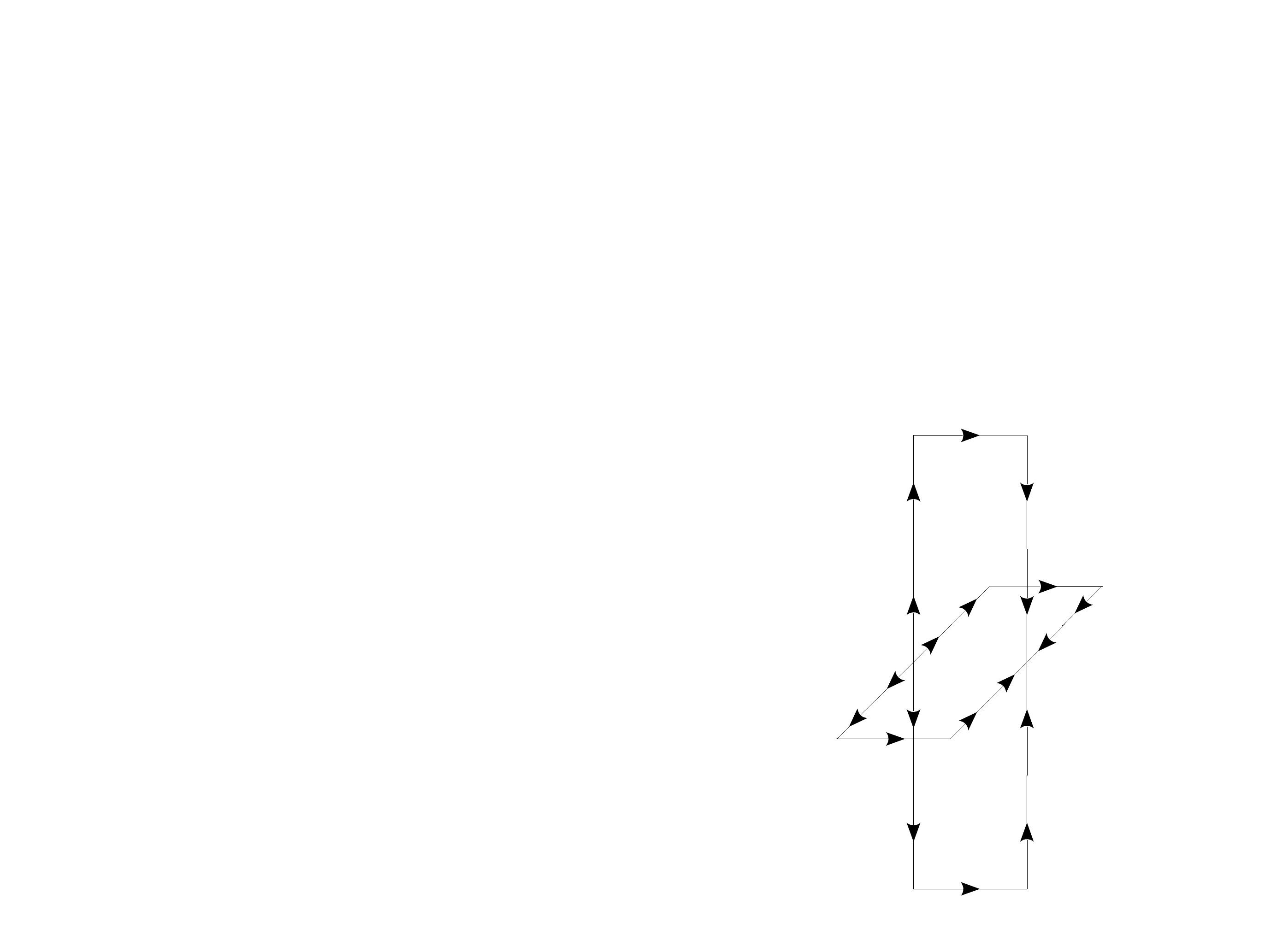}\tabularnewline
\end{tabular}
\par\end{centering}

\caption{Staples used in the extended spatial smearing. \label{fig:staples}}
\end{figure}

To increase the ground state overlap, we use a spatial extended smearing

\begin{equation}
U_{i}\rightarrow\mathcal{P}_{SU(3)}\Big[U_{i}+w_{1}\sum_{j}S_{ij}^{1}+w_{2}\sum_{j}S_{ij}^{2}+w_{3}\sum_{j}S_{ij}^{3}\Big]
\end{equation}

with staples $S_{ij}^{1}$, $S_{ij}^{2}$ and $S_{ij}^{3}$ being
given on figure \ref{fig:staples}. 

Even using this technique we weren't able to find a value of $t$
for which the plaquette to Wilson Loop correlators are stable within
error bars, while still have a sufficiently high signal to noise ratio.
To solve this, we note that the correlator which gives the average
of field $\langle F\rangle$ should be given by the formula $\langle F\rangle_{t}=\langle F\rangle_{\infty}+b\, e^{-\Delta t}$
for large values of $t$, with $\Delta=V_{1}-V_{0}$, being the difference
between the first excited and the ground state potentials. To compute
$\Delta$, we a use a variational basis of four levels of APE smearing,
with the potentials $V_{1}$ and $V_{0}$ being given by the solution
of the variational generalized eigensystem 
\begin{equation}
\langle W_{ij}(t)\rangle c_{j}^{n}(t)=w_{n}(t)\langle W_{ij}(0)\rangle c_{j}^{n}(t)\label{eq:variational}
\end{equation}

where $\langle W_{ij}\rangle=\langle\mathcal{O}_{i}(t)\mathcal{O}_{j}^{\dagger}(0)\rangle$
is the correlation between the meson annihilation and creation operators
at time $t$ and $0$ in the smeared states $i$ and $j$ respectively.

\subsection{Results and conclusions}

The results for the Lagrangian density in the quark-antiquark mediator
plane are shown on the left side of Fig. \ref{fig:meson_results}.
As can be seen the tube flux becomes wider as the quark-antiquark
distance is increased. To estimate the width of the flux tube we define
the value $w_{1/e}$ by $\langle\mathcal{L}\rangle(w_{1/e})=\frac{1}{e}\langle\mathcal{L}\rangle(0)$.
The results for this are shown on the right side of Fig. \ref{fig:meson_results}.

It was predicted \cite{Luscher:1980iy,Gliozzi:2010zv} that the squared width $w^{2}$
of this flux tube diverges logarithmicaly as $R\rightarrow\infty$,
that is $w^{2}\sim w_{0}^{2}\,\log\frac{R}{R_{0}}$ , where $w^{2}$.
This behavior is called ``roughening''. So, we can say that the
flux tube roughening has been observed from $R\sim0.4\, fm$ to $R\sim1.2\, fm$.
However, we weren't yet able to confirm the that the behavior is logarithmic.

\begin{center}
\begin{figure}
\begin{centering}
\includegraphics[width=0.4\textwidth]{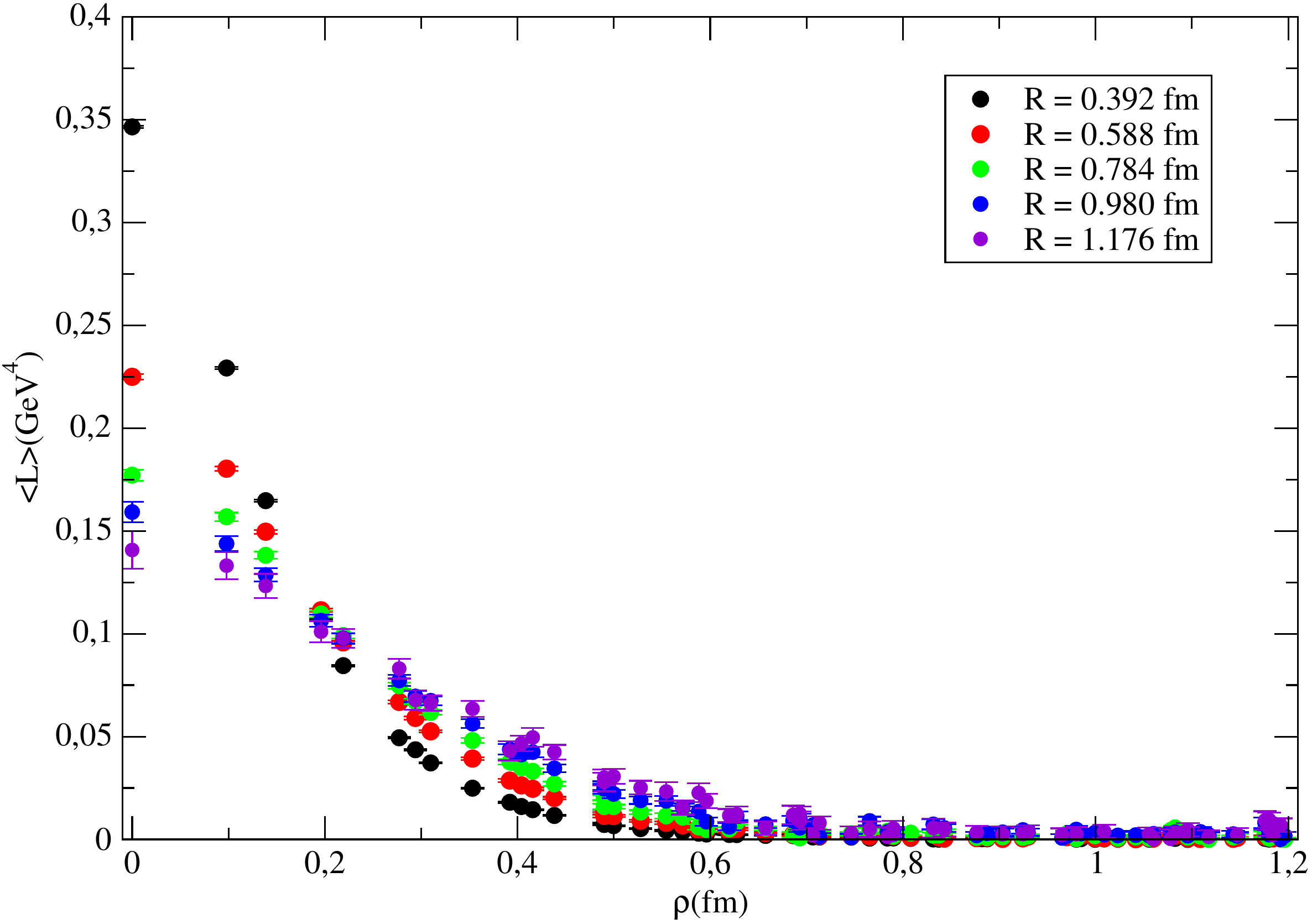}\hspace{1cm}\includegraphics[width=0.4\textwidth]{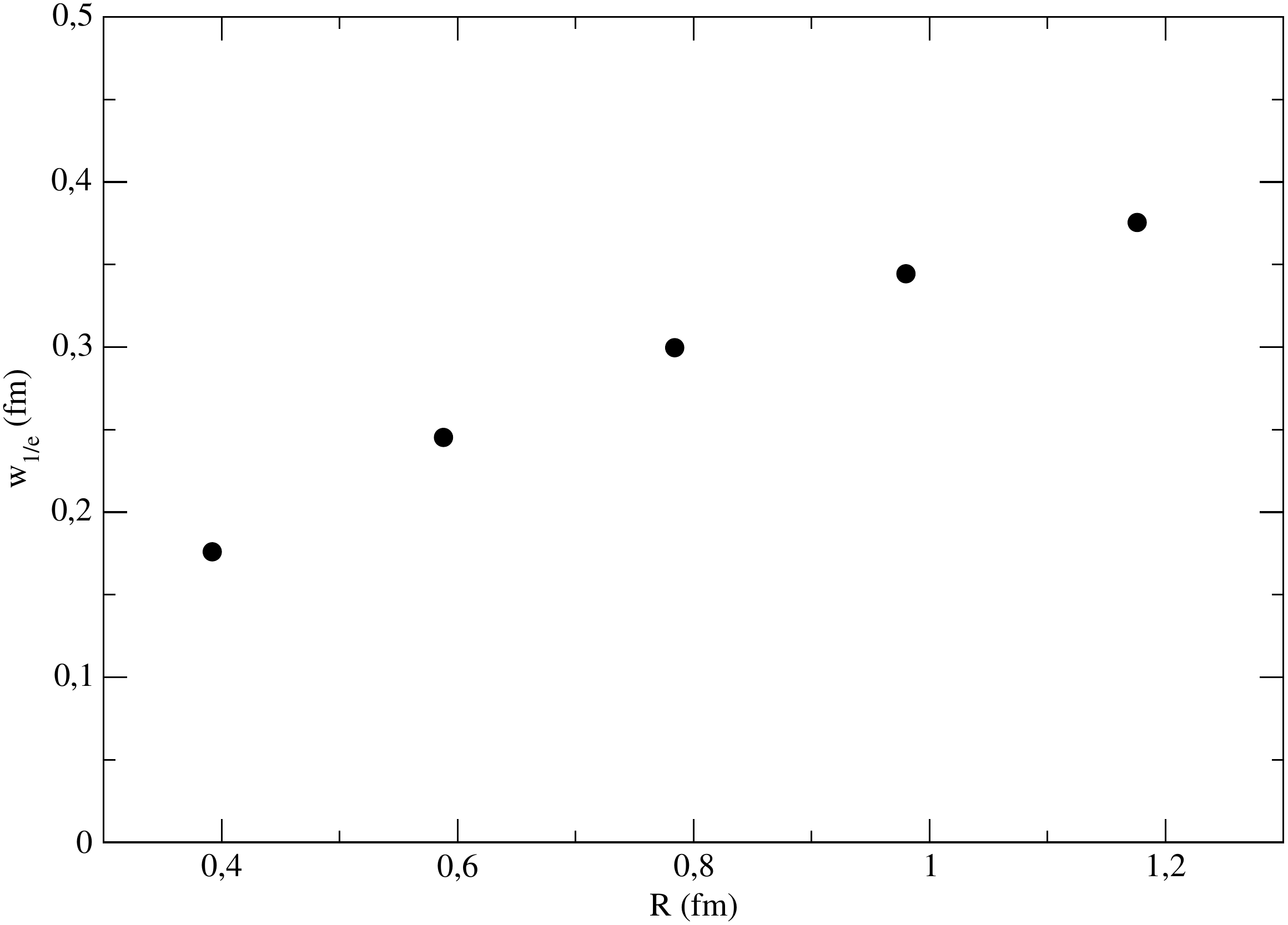}
\par\end{centering}

\caption{\label{fig:meson_results}Left: Lagrangian density in the mediator
plane between the quark and the antiquark. Right: Width of the flux
tube $w_{1/e}$ as a function of $R$.}
\end{figure}

\par\end{center}

\section{$QQ\bar{Q}\bar{Q}$ system}

Now, we consider a system of two quarks $Q_{1}\, Q_{2}$ and two antiquarks
$\overline{Q}_{3}\,\overline{Q_{4}}$. The study of this kind of systems
is important for the understanding of meson meson scattering processes
and the possible formation of tetraquarks, particles made of two valence
quarks and two valence antiquarks.

In this system we have two linearly independent color singlets. Those
can be, for instance, the two meson states: $|I\rangle=\frac{1}{3}|Q_{i}Q_{j}\overline{Q}_{i}\overline{Q}_{j}\rangle$
and $|II\rangle=\frac{1}{3}|Q_{i}Q_{j}\overline{Q}_{j}\overline{Q}_{i}\rangle$,
or the anti-symmetric and symmetric color states: $|A\rangle=\frac{\sqrt{3}}{2}\big(|I\rangle-|II\rangle\big)$
and $|S\rangle=\sqrt{\frac{3}{8}}\big(|I\rangle+|II\rangle\big)$.
In accordance with lattice results \cite{Alexandrou:2004ak,Okiharu:2004ve},
the ground state could be either $|I\rangle$, $|II\rangle$ or $|A\rangle$,
with the static potential being given by the flip-flop potential $V_{FF}=\min(V_{I},V_{II},V_{T})$,
where $V_{I}$ and $V_{II}$ are the two possible two-meson potentials,
given by the sum of the intra-meson potentials, and $V_{T}$ is the
tetraquark potential, which confines all the particles, with the confining
part being proportional to the minimal length of a fundamental string
linking the four particles.

\subsection{Geometries}

\begin{figure}
\begin{centering}
\includegraphics[width=0.25\textwidth]{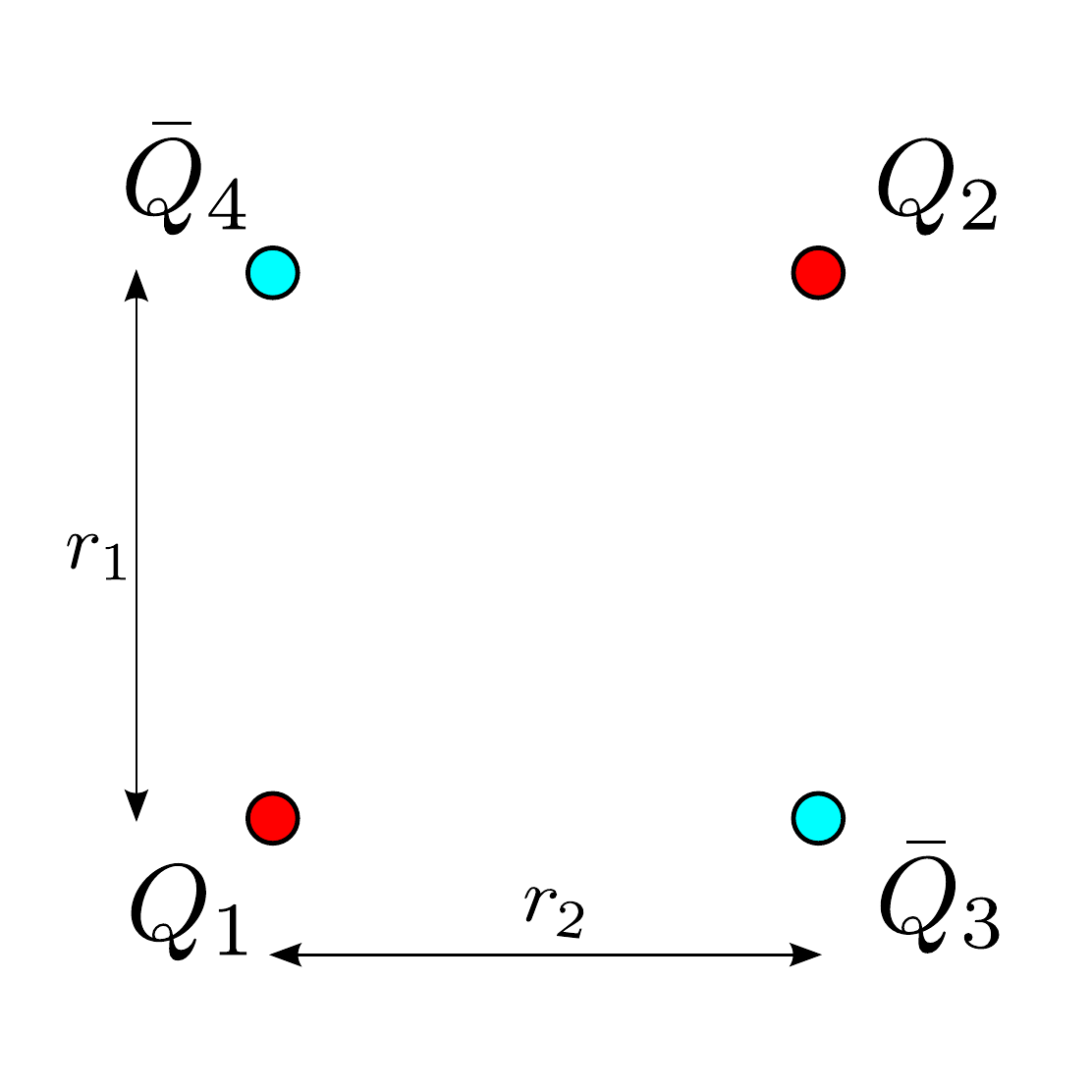}\hspace{1cm}\includegraphics[width=0.25\textwidth]{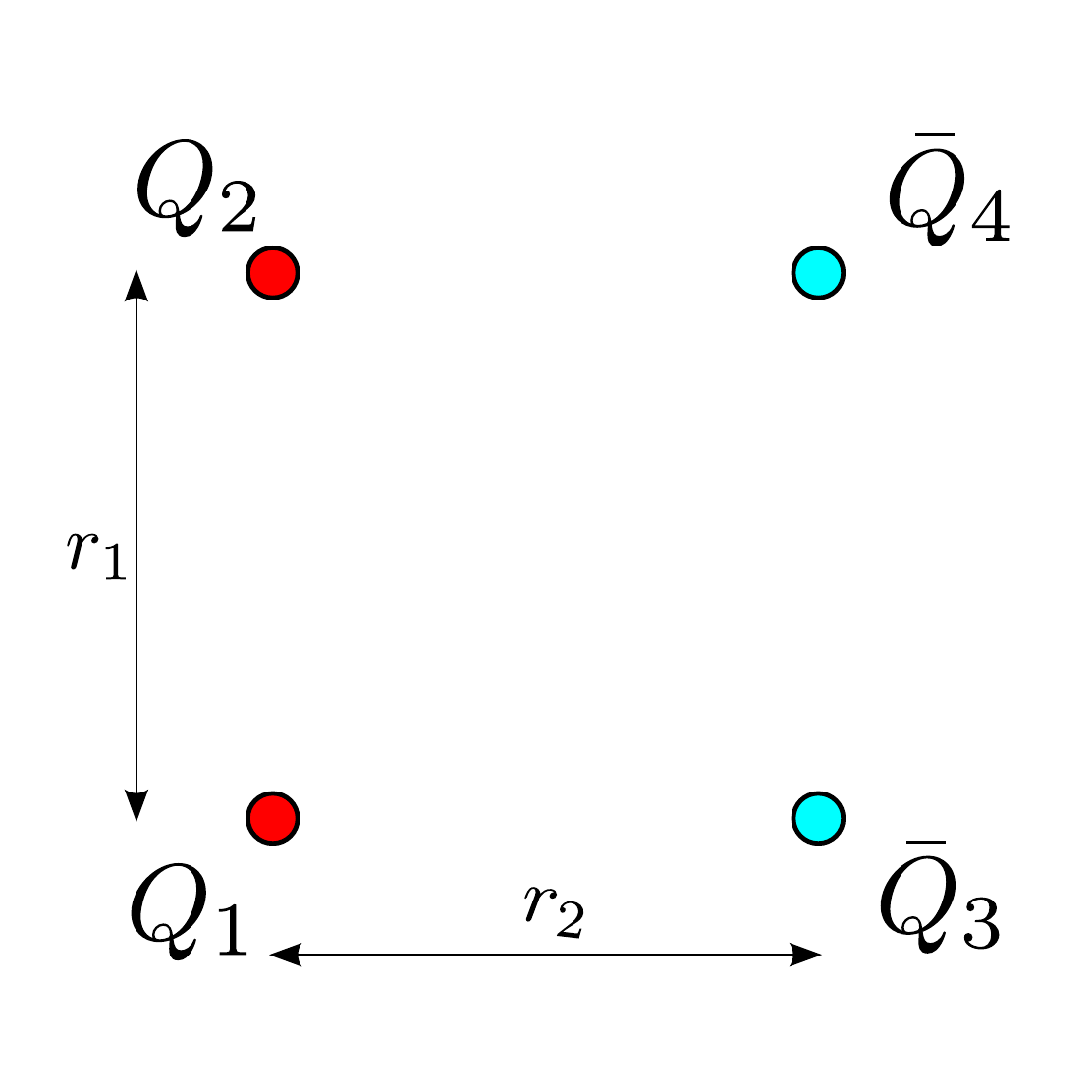}
\par\end{centering}

\caption{Left: Anti-parallel geometry. Right: Parallel geometry. \label{fig:geometries}}
\end{figure}

We use the two geometries, shown in Fig. \ref{fig:geometries}. In
both, the four particles form a rectangle, however in one of them
--- the parallel --- similar particles are on the same side of the
rectangle while on the other they are on opposite corners --- the
anti-parallel. With the first one, we can study the tetraquark to
mesons transition while with the second we can see the transition
between the two meson states.

In this work, we use 1121 Quenched $24^{3}\times48$ lattice configurations
with $\beta=6.2$. APE Smearing was applied to the spatial links,
while \foreignlanguage{american}{Hyper-cubic} blocking \cite{Hasenfratz:2001hp}
was applied to the temporal links.

\subsection{Variational method}

\begin{figure}
\begin{centering}
\includegraphics[width=0.4\textwidth]{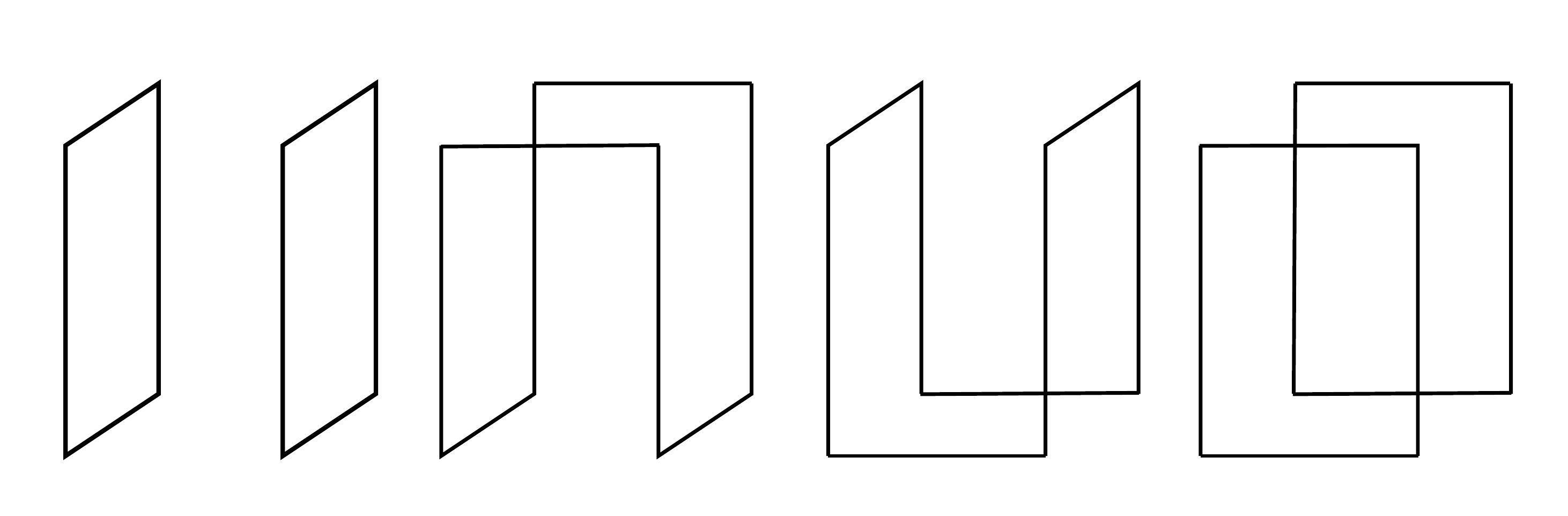}\hspace{1cm}\includegraphics[width=0.4\textwidth]{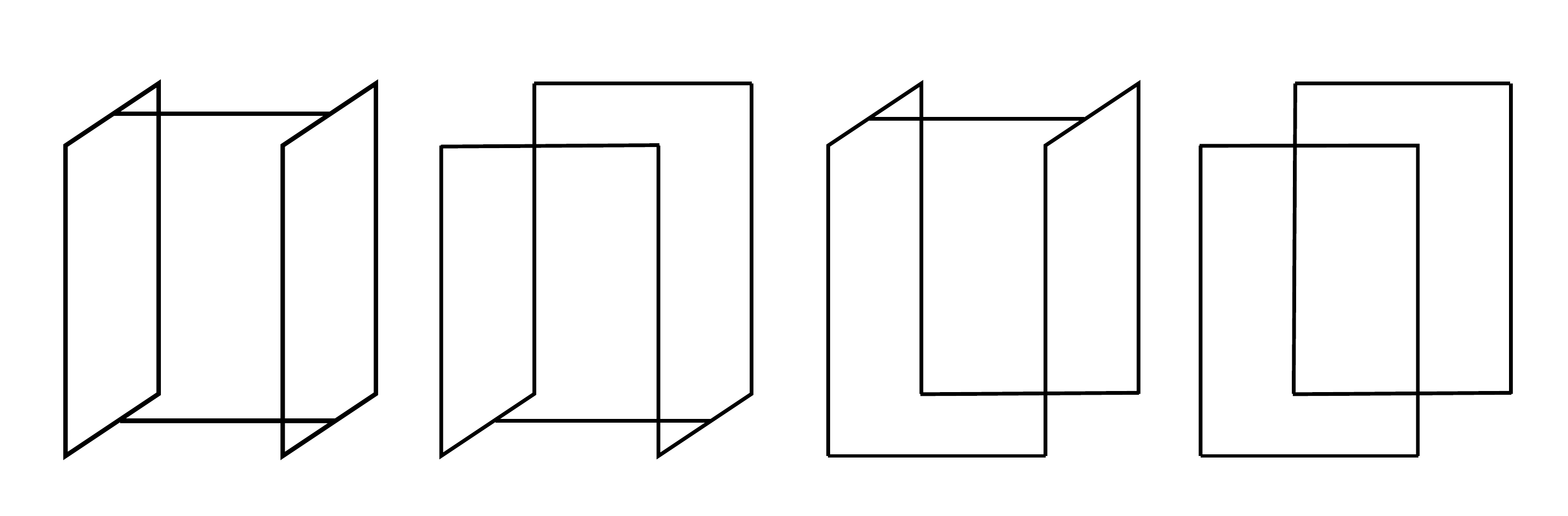}
\par\end{centering}

\caption{Left: Loops used to study the antiparallel geometry. Right: Loops
used to study the parallel geometry. \label{fig:loops}}
\end{figure}

To obtain not only the ground state, but also the first excited state,
we use a variational basis similar to Eq. (\ref{eq:variational}) but
using two different kind of operators. This way, the Wilson loop
which appears on Eqs. (\ref{eq:plaqE}) and (\ref{eq:plaqB}) is $W_{n}=c_{n}^{i}W_{ij}c_{n}^{j}$.
The two operator basis will be formed by the $|I\rangle$ and $|II\rangle$
annihilation operators in the anti-parallel geometry and in the parallel
geometry, by the $|I\rangle$ and $|A\rangle$ \cite{Alexandrou:2004ak,Okiharu:2004ve}
annihilation operators. This gives the matrix elements on Fig. \ref{fig:loops}.

\subsection{Results}

\begin{figure}
\begin{centering}
\includegraphics[bb=50bp 10bp 240bp 180bp,clip,width=0.25\textwidth]{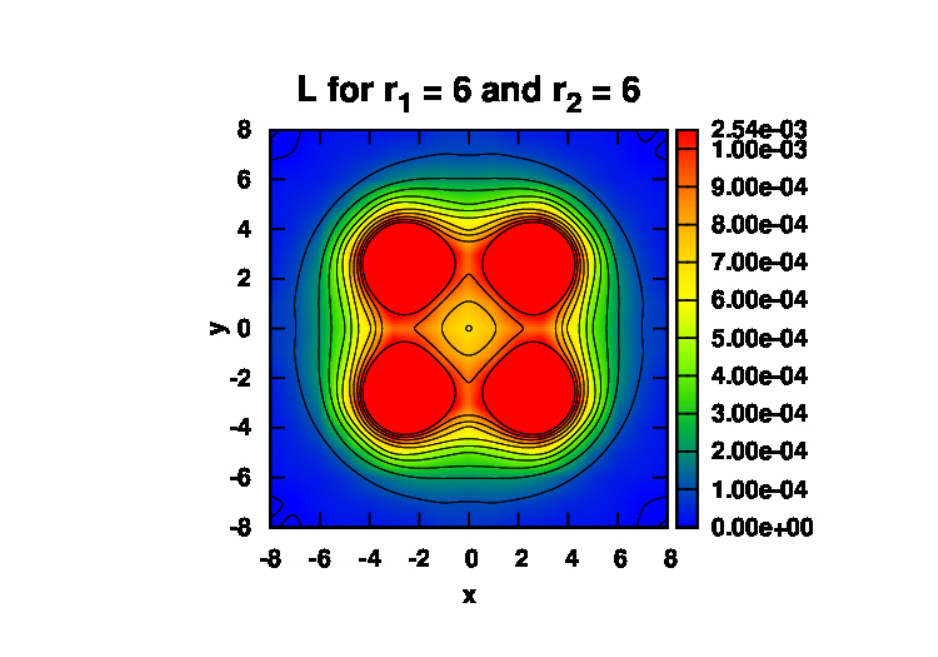}\includegraphics[bb=50bp 10bp 240bp 180bp,clip,width=0.25\textwidth]{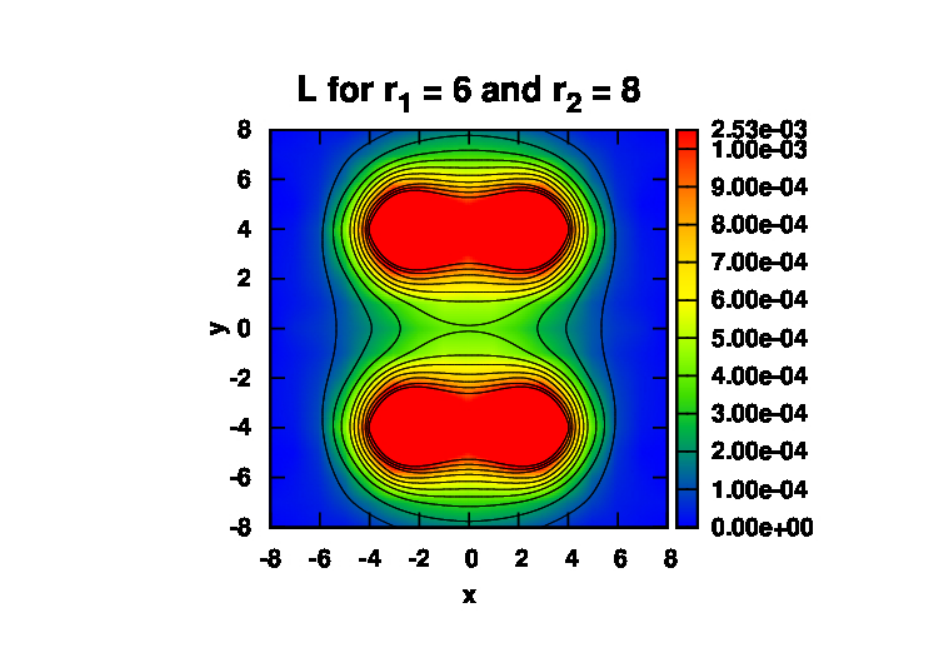}\includegraphics[bb=50bp 10bp 240bp 180bp,clip,width=0.25\textwidth]{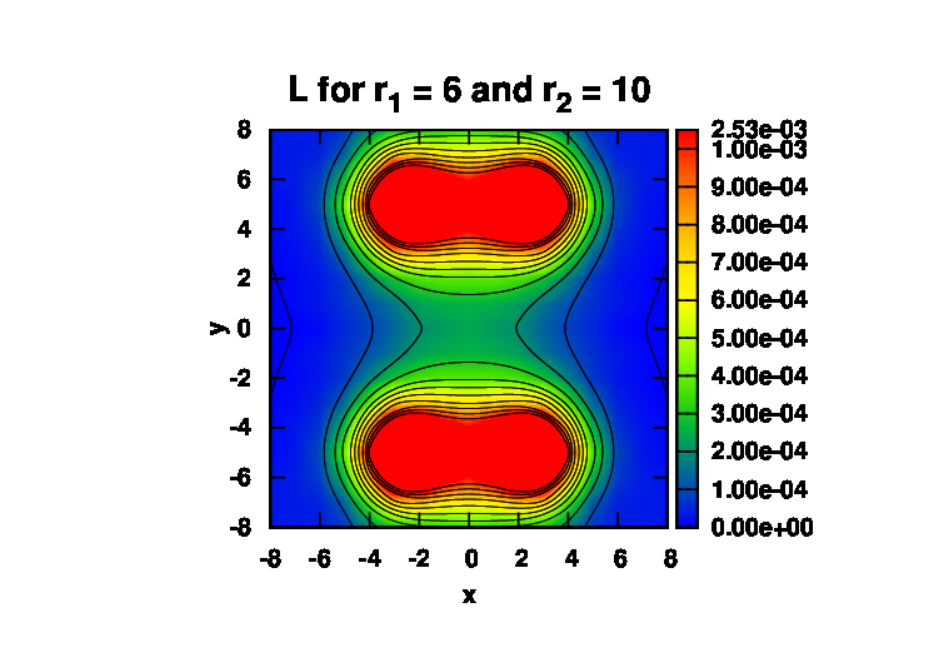}
\par\end{centering}

\caption{Lagrangian density for the ground state of the antiparallel geometry.
\label{fig:antip_n0}}
\end{figure}

\begin{figure}
\begin{centering}
\includegraphics[bb=50bp 10bp 240bp 180bp,clip,width=0.25\textwidth]{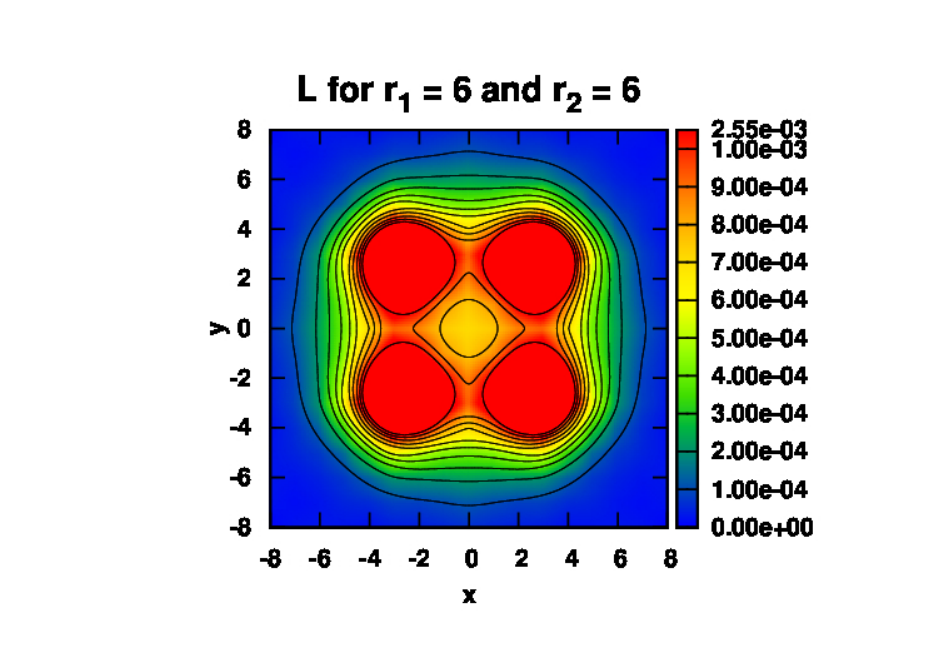}\includegraphics[bb=50bp 10bp 240bp 180bp,clip,width=0.25\textwidth]{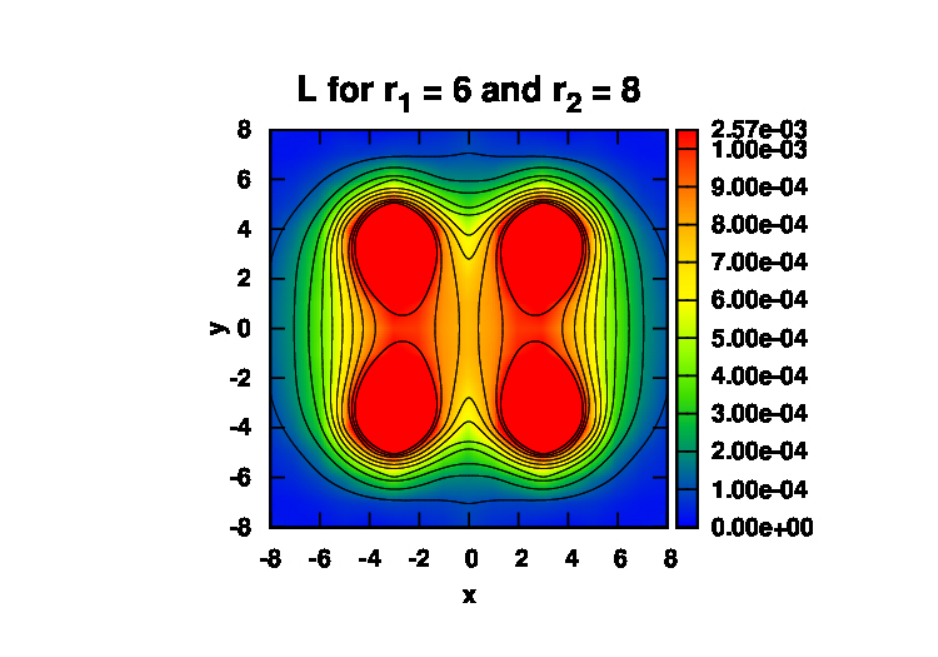}\includegraphics[bb=50bp 10bp 240bp 180bp,clip,width=0.25\textwidth]{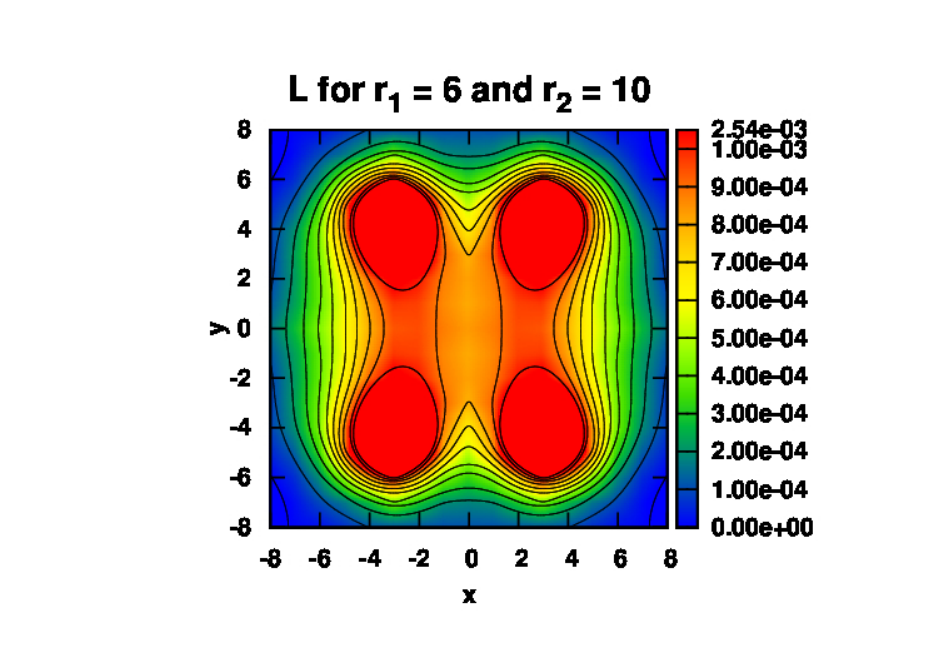}
\par\end{centering}

\caption{Lagrangian density for the first excited state of the antiparallel
geometry. \label{fig:antip_n1}}
\end{figure}

\begin{figure}
\begin{centering}
\includegraphics[bb=50bp 10bp 240bp 180bp,clip,width=0.25\textwidth]{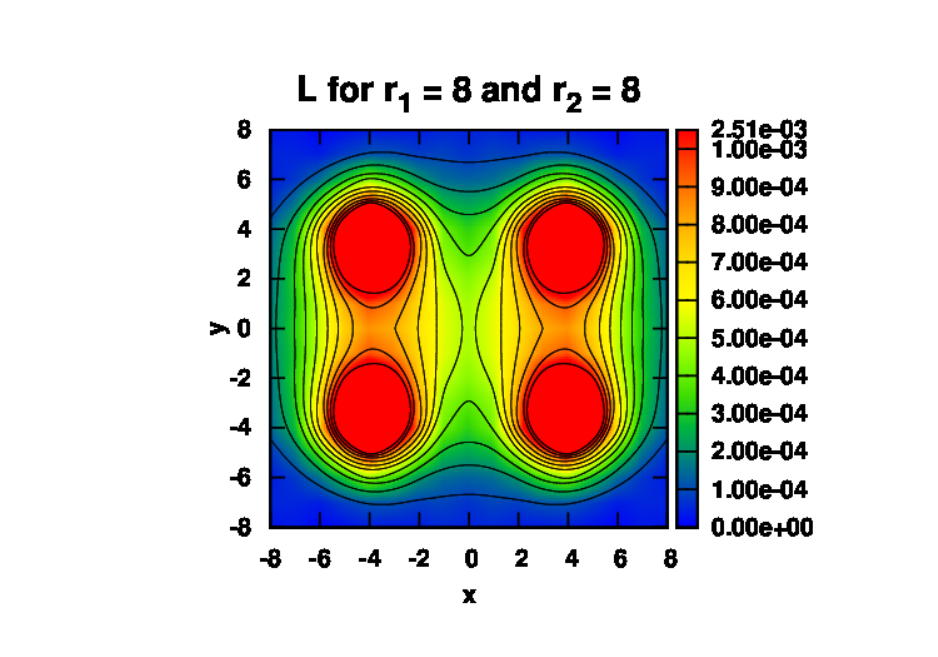}\includegraphics[bb=50bp 10bp 240bp 180bp,clip,width=0.25\textwidth]{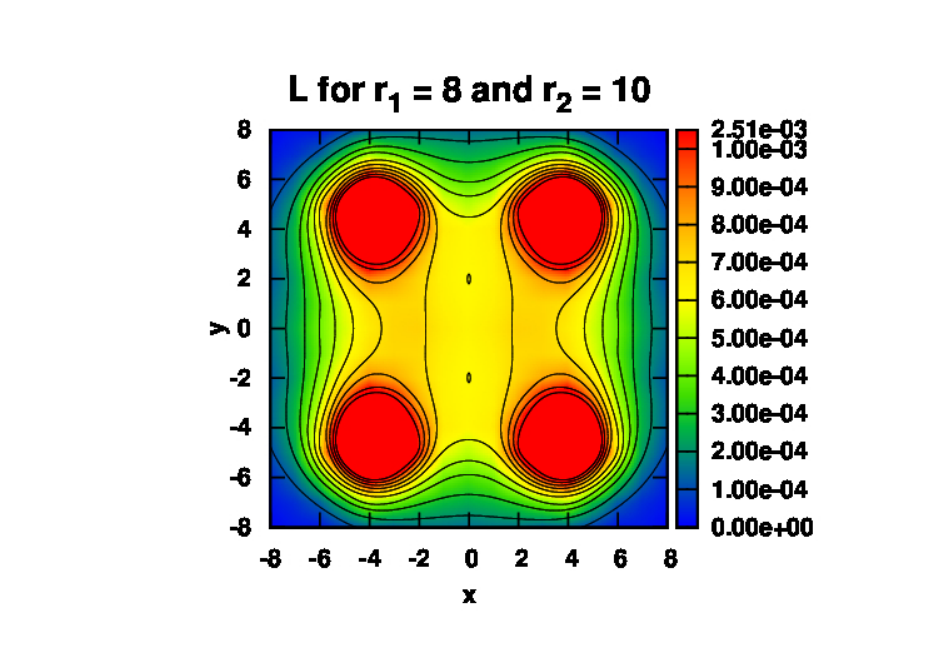}\includegraphics[bb=50bp 10bp 240bp 180bp,clip,width=0.25\textwidth]{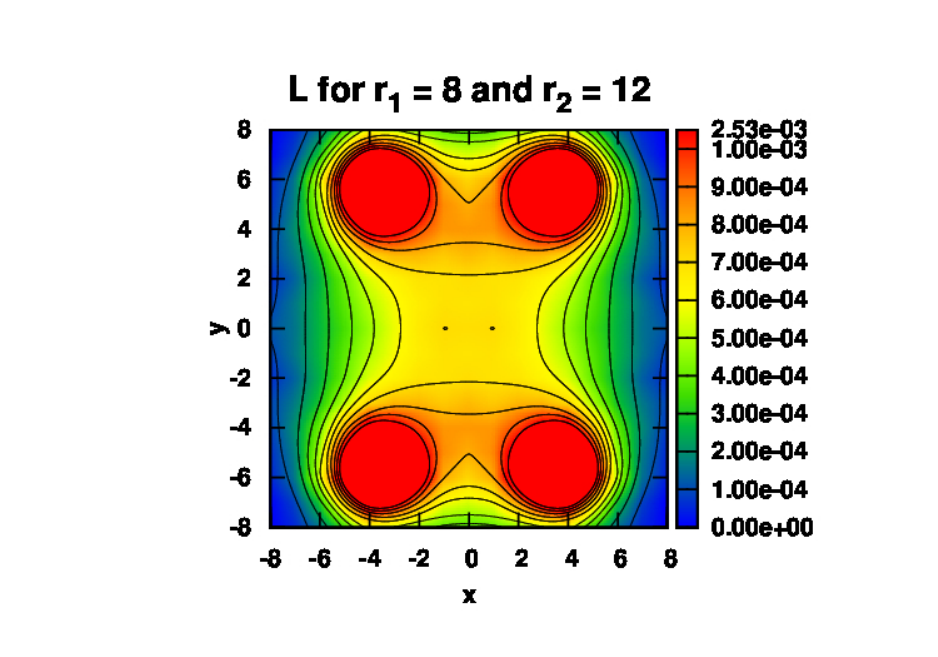}\includegraphics[bb=50bp 10bp 240bp 180bp,clip,width=0.25\textwidth]{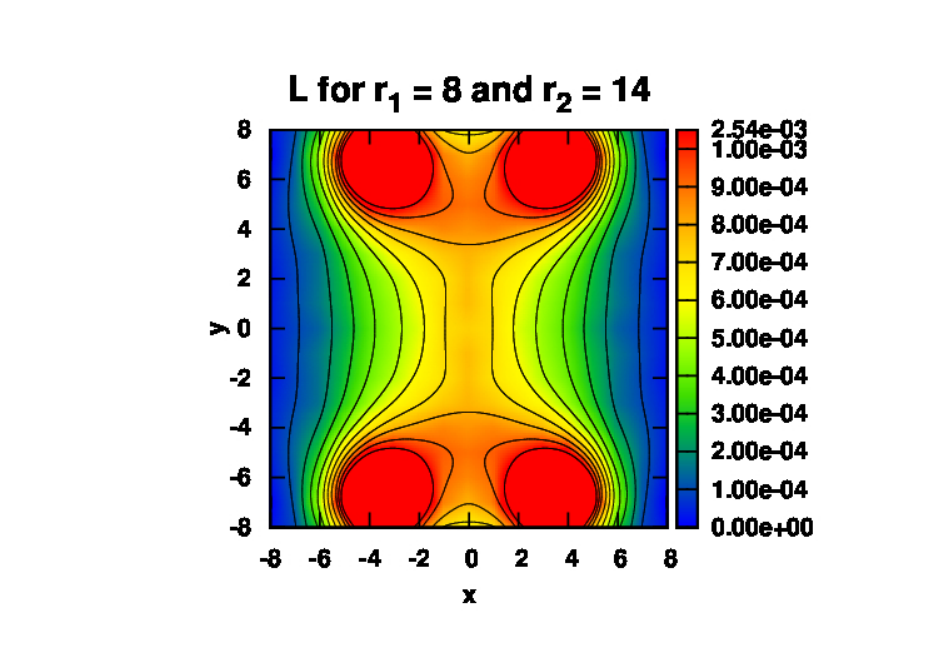}
\par\end{centering}

\caption{Lagrangian density for the ground state of the parallel geometry.
\label{fig:para_n0}}
\end{figure}

The results for the Lagrangian density $\mathcal{L}$ in the ground
state and the first excited state are given on Figs. \ref{fig:antip_n0}
and \ref{fig:antip_n1}. For the ground state the system collapses
in a two meson state, when $r_{2}\gg r_{1}$, as expected. Looking
at the Wilson loop composition we also conclude that when $r_{1}=r_{2}$
the ground state is color symmetric. The excited state is not so readily
explainable. The results for $\mathcal{L}$ in the parallel geometry
can be seen on Figs. \ref{fig:para_n0} and \ref{fig:para_n1}. The
results are the expected ones for the ground state, with the system
passing from a two meson state to a tetraquark state as we increase
$r_{2}$. We discuss the first excited state in the conclusion.

\begin{center}
\begin{figure}
\begin{centering}
\includegraphics[bb=50bp 10bp 240bp 180bp,clip,width=0.25\textwidth]{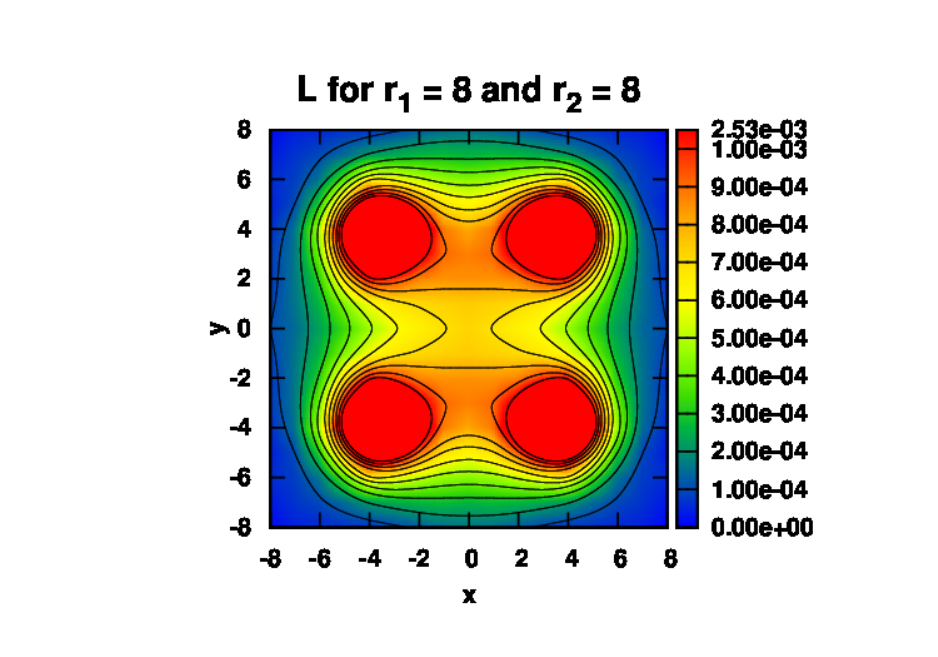}\includegraphics[bb=50bp 10bp 240bp 180bp,clip,width=0.25\textwidth]{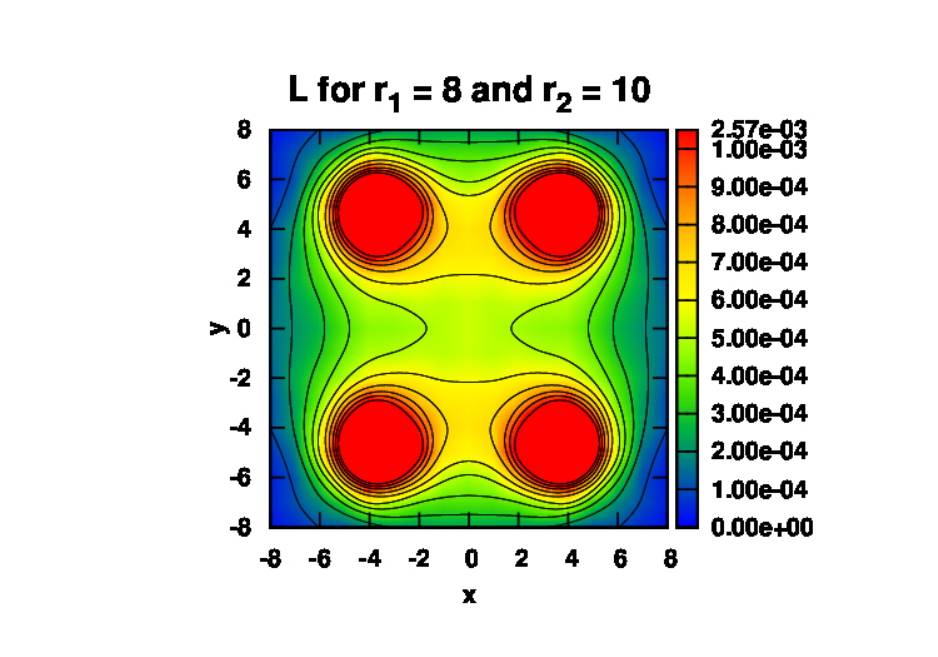}\includegraphics[bb=50bp 10bp 240bp 180bp,clip,width=0.25\textwidth]{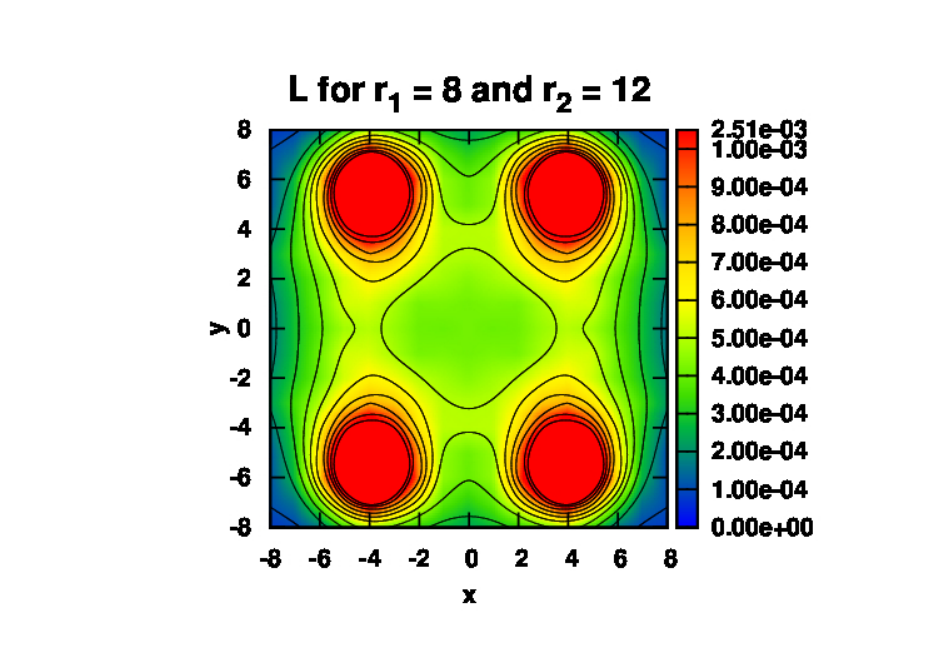}\includegraphics[bb=50bp 10bp 240bp 180bp,clip,width=0.25\textwidth]{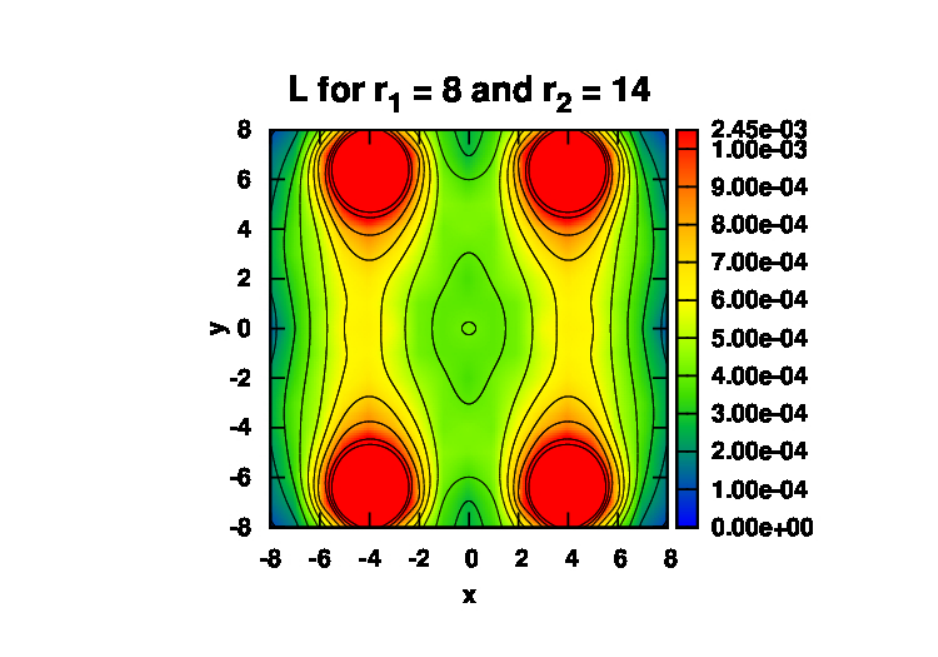}
\par\end{centering}

\caption{Lagrangian density for the first excited state of the parallel geometry.
\label{fig:para_n1}}
\end{figure}

\par\end{center}

\subsection{Discussion/Conclusion}

The results for the ground state are consistent with the flip-flop
ground state potential with the color fields dispositions being consistent
with $|I\rangle$, $|II\rangle$ or $|A\rangle$ where expected.

We can explain the excited states if we consider that the vibrational and rotational
flux tube excitations effects are negligible. In this case the first excited
state is orthogonal to the ground state. In that case, if we compute
the Casimir factors for this orthogonal states, as in \cite{Cardoso:2012uk},
that they predict the correct behavior of the fields, with a repulsion
between particles, where we expect it.

\bibliographystyle{plain}
\bibliography{bib}

\end{document}